\documentclass[aps,prb,twocolumn,twoside,superscriptaddress]{revtex4}

\usepackage[version=3]{mhchem} 
\usepackage{physics}
\usepackage{graphicx}
\usepackage{amsmath}
\usepackage{float}
\usepackage{amssymb}
\usepackage{placeins}
\usepackage{array}
\usepackage{subfig}
\usepackage{color}
\usepackage{makecell}
\usepackage{mhchem}
\usepackage{siunitx}
\usepackage{threeparttable}
\usepackage{xr}
\usepackage{xstring}
\usepackage{upgreek}
\usepackage{caption}

\newcommand{\comment}[1]{}

\newcommand{\vecauto}[1]{\ensuremath{\vectorbold*{#1}}}

\newcommand{\vR}{\vecauto{R}}

\newcommand{\CN}{\ensuremath{\textrm{C}_\textrm{N}}}
\newcommand{\NBVN}{\ensuremath{\textrm{N}_\textrm{B}\textrm{V}_\textrm{N}}}
\newcommand{\CBVN}{\ensuremath{\textrm{C}_\textrm{B}\textrm{V}_\textrm{N}}}
\newcommand{\XBVN}{\ensuremath{\textrm{X}_\textrm{B}\textrm{V}_\textrm{N}}}
\newcommand{\OBOBVN}{\ensuremath{\textrm{O}_\textrm{B}\textrm{O}_\textrm{B}\textrm{V}_\textrm{N}}}

\newcommand{\Bandname}[1]{\ensuremath{
  \StrChar{#1}{1}[\chara]
  \StrChar{#1}{2}[\charb]
  \StrChar{#1}{3}[\charc]
  \StrChar{#1}{4}[\chard]
\IfEqCase{\chard}{
   {U}{\chara\mathrm{\charb}_{\charc}\uparrow}
   {D}{\chara\mathrm{\charb}_{\charc}\downarrow}
}[\PackageError{Bandname}{Undefined option for spin: \StrChar{#1}{4}}]
}}

\definecolor{tcolor}{rgb}{0, 0.5, 0.2}

\begin{document}

\title{Carrier Recombination Mechanism at Defects in Wide Bandgap Two-dimensional Materials from First-principles}

\author{Feng Wu\footnote[2]{FW and TJS contributed equally to this work.}}
\affiliation{Department of Chemistry and Biochemistry, University of California Santa Cruz, Santa Cruz, CA, 95064, USA}
\author{Tyler J. Smart\footnotemark[2]}
\affiliation{Department of Physics, University of California Santa Cruz, Santa Cruz, CA, 95064, USA}
\author{Junqing Xu}
\affiliation{Department of Chemistry and Biochemistry, University of California Santa Cruz, Santa Cruz, CA, 95064, USA}
\author{Yuan Ping\footnote[1]{yuanping@ucsc.edu}}
\affiliation{Department of Chemistry and Biochemistry, University of California Santa Cruz, Santa Cruz, CA, 95064, USA}


\begin{abstract}
Identification and design of defects in two-dimensional (2D) materials as promising single photon emitters (SPE) requires a deep understanding of underlying carrier recombination mechanisms. 
Yet, the dominant mechanism of carrier recombination at defects in 2D materials has not been well understood, and some outstanding questions remain: How do recombination processes at defects differ between 2D and 3D systems? What factors determine defects in 2D materials as excellent SPE at room temperature?
In order to address these questions, we developed first-principles methods to accurately calculate the radiative and non-radiative recombination rates at defects in 2D materials, using $h$-BN as a prototypical example. We reveal the carrier recombination mechanism at defects in 2D materials being mostly dominated by defect-defect state recombination in contrast to defect-bulk state recombination in most 3D semiconductors. 
In particular, we disentangle the non-radiative recombination mechanism into key physical quantities: zero-phonon line (ZPL) and Huang-Rhys factor. 
At the end, we identified strain can effectively tune the electron-phonon coupling at defect centers and drastically change non-radiative recombination rates.
Our theoretical development serves as a general platform for understanding carrier recombination at defects in 2D materials, while providing pathways for engineering of quantum efficiency of SPE.

\end{abstract}

\maketitle

The engineering of spin defects in wide-band semiconductors offers a promising avenue for the development of quantum spin devices. \cite{Aharonovich2011, Ladd2010, Seo2016, Dreyer2018} They are among the few alternatives for quantum technologies that operate at room temperature. Deep defects in two-dimensional (2D) materials such as hexagonal boron nitride ($h$-BN) \cite{Aharonovich2014,Tran2016,Kim2018,Bourrellier2016,Grosso2017,Li2017,Exarhos2017,Exarhos2019,Museur2008,Sontheimer2017,Jungwirth2016,Jungwirth2017,Noh2018} and transition metal dichalcogenides (TMD) \cite{Mak2016,Palacios-Berraquero2017} have proven to be promising single photon sources with polarized and ultrabright single photon emission at room temperature. These materials exhibit unprecedented potential for several applications, including large-scale nanophotonics and quantum information processing,~\cite{Beveratos2002,OBrien2009,Scarani2009,Aharonovich2016,Vogl2018} which in turn provide a new platform for exploring quantum phenomena. \cite{Dreyer2018} 
In order for these defect centers to provide bright SPE,~\cite{Aharonovich2011,Aharonovich2016} the radiative recombination rate (photon emitted) needs to be high, while the non-radiative recombination rate (no photon emitted) must be substantially lower to yield a high quantum efficiency. 
Furthermore, a weak electron-phonon coupling is also required to ensure long spin relaxation time for the application of qubit and stable single photon emission at room temperature.

Despite the importance of maximizing radiative rates for quantum information, the factors which determine the recombination process at defects in 2D materials are not understood experimentally or theoretically. Past theoretical studies have either focused on radiative recombination in pristine 2D materials \cite{Palummo2015,Wu2017} or phonon assisted non-radiative recombination for defects in 3D wide band-gap semiconductors.~\cite{Alkauskas2014,Shi2015} Therefore, a fully comparative study of both recombination processes for defect centers in 2D materials is highly desired. 

\begin{figure}[t]
  \includegraphics[width=0.8\linewidth]{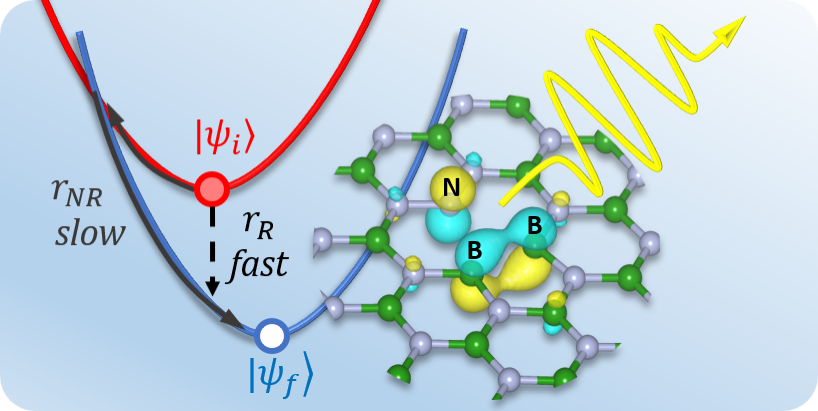}
  \caption{Schematic diagram of carrier recombination at the \NBVN\ defect in monolayer $h$-BN. In order for a defect to be a robust single photon source, it is necessary to have the radiative recombination rates much higher than the non-radiative ones.}
  \label{fig:toc1}
\end{figure}

Furthermore, the high tunability of SPE allows them to be integrated within a vast array of applications. \cite{Tran2016,Aharonovich2016} Among these methods, strain modulation is one of the most effective strategies, especially for low-dimensional materials which can work under large distortion.~\cite{Brawand2015}
For example, in 2D systems, some key electronic properties such as band gap, change by 1.5\% under 1\% uniaxial tension in TMD monolayers \cite{Chang2013} or 6\% under 1\% uniaxial tension in phosphorene\cite{Peng2014}. Additionally, the non-radiative process, which is intrinsically sensitive to lattice deformation (as it is phonon mediated), may exhibit even more drastic changes under strain. 

In this communication, we first introduce the formalism of computing the radiative and non-radiative lifetime of defect excited states from first-principles. We then focus on comparing radiative and non-radiative processes of different transitions in a series of important defects in monolayer $h$-BN, where we discuss the dominant recombination processes and their implication on SPE efficiency. 
Finally, we show that applying strain to $h$-BN defects can effectively tune the non-radiative rates and quantum yield of SPE.

The radiative and non-radiative transition rates between two electronic states under perturbation can be computed via Fermi's golden rule:
\begin{align}
    r_{if}^{R} &=  \frac{2\pi}{\hbar}g\left|\mel{f}{H^R}{i}\right|^2\delta(E_i-E_f),  \label{eq:rate-radiative-fg} \\
    r_{if}^{NR} &= \frac{2\pi}{\hbar}g\sum_{n,m}p_{in}\left|\mel{fm}{H^{e-ph}}{in}\right|^2\delta(E_{in}-E_{fm}). \label{eq:rate-nonradiative-fg}
\end{align}
Here, $r_{if}^R$ and $r_{if}^{NR}$ denote the recombination rates between electronic states $i$ and $f$ via a radiative process ($r_{if}^R$) and non-radiative process ($r_{if}^{NR}$), respectively. 
$g$ is the degeneracy factor of the final state, i.e. several equivalent energy-degenerate atomic configurations of the final state might exist.~\cite{Alkauskas2014a}
For defects in 2D materials studied in this work, $g$ factors are all equal to 1.
$H^R$ is the electron-photon coupling (electromagnetic) Hamiltonian and $H^{e-ph}$ is the electron-phonon coupling Hamiltonian. A sum over phonon states $n,m$ enters the non-radiative recombination process with an occupation number $p_{in}$ of the vibronic state $\ket{in}$. For ground state calculations, we employed open source plane-wave code Quantum ESPRESSO~\cite{QE1} with ONCV norm conserving pseudopotentials~\cite{ONCV1,ONCV2} and a supercell size of $6\times6$ or higher. Charge corrections for the total energies and eigenvalues of charged defects were applied by employing the techniques developed in Ref.~\citenum{Wu2017, PingJCP}. The total energies, defect formation energies and geometry were computed at both PBE and hybrid functional levels (the results presented in the main text are computed at PBE, and detailed comparison between two levels can be found in the Supporting Information (SI) Table S4). The band gaps of pristine $h$-BN are computed at $GW@PBE$ as done in our previous work~\cite{Wu2017}, which are 6.01 eV for bulk and 7.01 eV for monolayer $h$-BN respectively. 
The exciton dipole moments and exciton energies as input for radiative lifetime were computed at many body perturbation theory with GW approximation for quasiparticle energies~\cite{WEST,Pingreview,Viet2012} and then solving the Bethe-Salpeter equation with the Yambo-code~\cite{YAMBO}, as well as Random Phase approximation with DFT eigenvalues (detailed comparison can be found in the SI, Table S3; the results in the main text are computed at DFT-RPA). More computational details and formulation of radiative rates $r_{if}^R$ are discussed in the SI and Ref.~\citenum{Wu2019}. 

The non-radiative rate is simplified by the static coupling approximation with a one-dimensional (1D) effective phonon approximation\cite{Alkauskas2014,Alkauskas2014a,Barmparis2015,Alkauskas2012,Alkauskas2016,Brawand2015,Giustino2017,Henry1977,Howgate1969,Monserrat2018,Passler1974,Passler1976,Shi2012,Shi2015,Wickramaratne2018,Shi2018,Yang2016,Book_TheoryofDefects_Stoneham} (the validation of 1D effective phonon approximation in $h$-BN is based on the similarity of Huang-Rhys factors between 1D effective phonon and all phonon calculations, as discussed in the SI, section IV):
\begin{align}
r_{if}^{NR} =& \frac{2\pi}{\hbar}g|W_{if}|^2 X_{if}(T), \label{eq:rate-nonrad-allterms}\\
X_{if}(T)=& \sum_{n,m}p_{in}\left|\mel{\phi_{fm}(\vR)}{Q-Q_a}{\phi_{in}(\vR)}\right|^2 \nonumber \\
& \hspace{1.5cm} \cross \delta(m\hbar\omega_f - n\hbar\omega_i+\Delta E_{if}), \label{eq:rate-nonrad-X} \\
W_{if} =& \mel{\psi_i(\mathbf{r},\vR)}{\frac{\partial H}{\partial Q}}{\psi_f(\mathbf{r},\vR)}\bigg{|}_{\vR=\vR_a},
\end{align}
where $r_{if}^{NR}$ is naturally separated into an electronic term $W_{if}$ and a phonon term $X_{if}(T)$ with temperature dependence from thermal population ($p_{in}$). Here $\Delta E_{if}$ is the zero phonon line energy (ZPL), which can be measured experimentally by photoluminescence. We implemented the non-radiative recombination rates as postprocessing codes of Quantum ESPRESSO~\cite{QE1}. 



A single defect may introduce several energy levels within the band gap of the host material. This yields the possibility for transitions to occur between defect states (``defect-defect'' transition), as well as from a defect state to a band edge (``defect-band'' transition). The computed non-radiative lifetimes and capture coefficients of the most probable defect-band transitions for hole captures in multiple defects in monolayer and bulk $h$-BN as well as bulk GaN are listed in Table~\ref{table:defect-band-lifetime} (where \XBVN\ (X=C, O, N) denotes X substitution of boron accompanied by a nitrogen vacancy). The capture coefficients are defined as a product of recombination rates $r_{ij}$ with surface area or volume for 2D or 3D systems, respectively. The corresponding defect formation energies and configuration coordinate diagrams are presented in the SI, Figure S1.
We find that all defect-band transitions in monolayer $h$-BN have very small rates (the corresponding lifetime exceeds milliseconds). This is in contrast to typical 3D bulk defects in other materials, such as GaN-\CN\ or GaN-$(\mathrm{Zn_{Ga}V_N})$, where the non-radiative lifetime is at the picosecond level with a similar defect concentration to $h$-BN. \cite{Shi2012,Alkauskas2014a,Reshchikov2014}

\begin{table}
\centering
    \caption{Non-radiative lifetimes and capture coefficients of defects in $h$-BN and GaN through defect-band recombination(only for the hole capture processes $A^{-1}+h^{+}\rightarrow A^{0}$).  For comparison a dominant defect-defect recombination at \NBVN\ in monolayer $h$-BN is also listed. The capture coefficients $C_p$ (with a unit of cm$^2/$s for 2D and cm$^3/$s for 3D systems) and lifetimes are reported at $T=300$ K. Lifetimes are defined as the inverse of rates $\uptau^{NR}=1/r_{if}^{NR}$ and computed in $6\times 6$ $h$-BN supercell or $2\times 2\times 2$ GaN supercell. 
   \label{table:defect-band-lifetime}}
  \begin{threeparttable}
    \begin{tabular}{cccccc}
      \hline
        System & \thead{ZPL \\ (eV)} & \thead{$\Delta Q$ \\ (amu$^{1/2}$\AA) }  & 
        \thead{$\hbar\omega_f$ \\ (meV) } &
        \thead{$C_p$ \\ (cm$^n$/s) } & $\uptau^{NR}$ \\
      \hline  
      ML \CBVN (2D) & 5.78 & 0.58 & 86 & $10^{-32}$ & $>1$ ms \\
      ML \OBOBVN (2D) & 4.26 & 0.84 & 85 &
      $10^{-29}$ & $>1$ ms \\
      ML \NBVN (2D) & 5.46 & 0.51 & 95 & $10^{-33}$ & $>1$ ms \\
      ML \CN (2D) & 3.87 & 0.35 & 150 & $10^{-16}$ & $>1$ ms \\
      Bulk \CN (3D) & 2.69 & 0.35 & 149 & $10^{-16}$ & 6.6 $\mu$s \\
      GaN-\CN (3D) & 1.00 & 1.39 & 39 & $10^{-9}$* & 0.29 ps \\ 
      \hline
      ML \NBVN (2D) & 2.04 & 0.53 & 100 & $10^{-4}$ & 102 ps \\
      (defect-defect) &&&&&\\
      \hline
    \end{tabular}
    \begin{tablenotes}
      \item * $7\times10^{-10}$ in Ref. \citenum{Alkauskas2014a}.
    \end{tablenotes}
\end{threeparttable}
\end{table}

One key reason that non-radiative defect-band recombination in monolayer $h$-BN is typically slow, is due to large energy differences between defect states and band edges ($\sim4-6$ eV) as the ZPL shown in Table~\ref{table:defect-band-lifetime}.~\cite{Wu2017,Smart2018,Rasmussen2016} Nonetheless, other factors such as the effective phonon frequencies $\hbar\omega_f$ and the change of nuclear positions $\Delta Q$ can also affect the rates, as discussed later.
For example, comparing monolayer BN-\CN\ with bulk BN-\CN, only the ZPL changes significantly (over 1 eV) and other parameters ($\hbar\omega_f$ and $\Delta Q$) retain nearly constants, which ultimately leads to a two order of magnitude difference in their capture rates.
Physically, the phonon-assisted non-radiative rate is dominated by a charge transfer process between initial and final state potential energy surfaces, and can be approximated by a classical Marcus' theory picture (see Figure \ref{fig:toc1}). Given the form of the energy barrier for charge transfer\cite{Brawand2015}, a large energy difference between the two states (ZPL) results in an exponential drop in the transfer rate (although exceptions can be found \cite{Nan2009,Jortner1976,Passler1982,Shi2015}). Therefore, in monolayer $h$-BN, the large ZPLs of defect-band transitions result in extremely slow non-radiative recombination processes (over milliseconds). 
On the other hand, several defects have allowed defect-defect transitions with viable non-radiative rates due to smaller energy differences, e.g. the \NBVN\ (nitrogen substitution of boron accompanied by a nitrogen vacancy) defect-defect transition in monolayer $h$-BN (Table \ref{table:defect-defect-NBVN}). 

Hence, the remaining discussions are focused on defect-defect transitions in monolayer $h$-BN. Defect-defect non-radiative recombination is performed for neutral excited and ground state with constrained occupation number. The equilibrium geometry, ZPL and vibrational frequency can be also obtained at DFT with constrained occupation. More computational details for defect-defect nonradiative recombination can be found in SI, section III, Figure S2 and Table S4. 


\begin{figure}
  \includegraphics[width=0.8\linewidth]{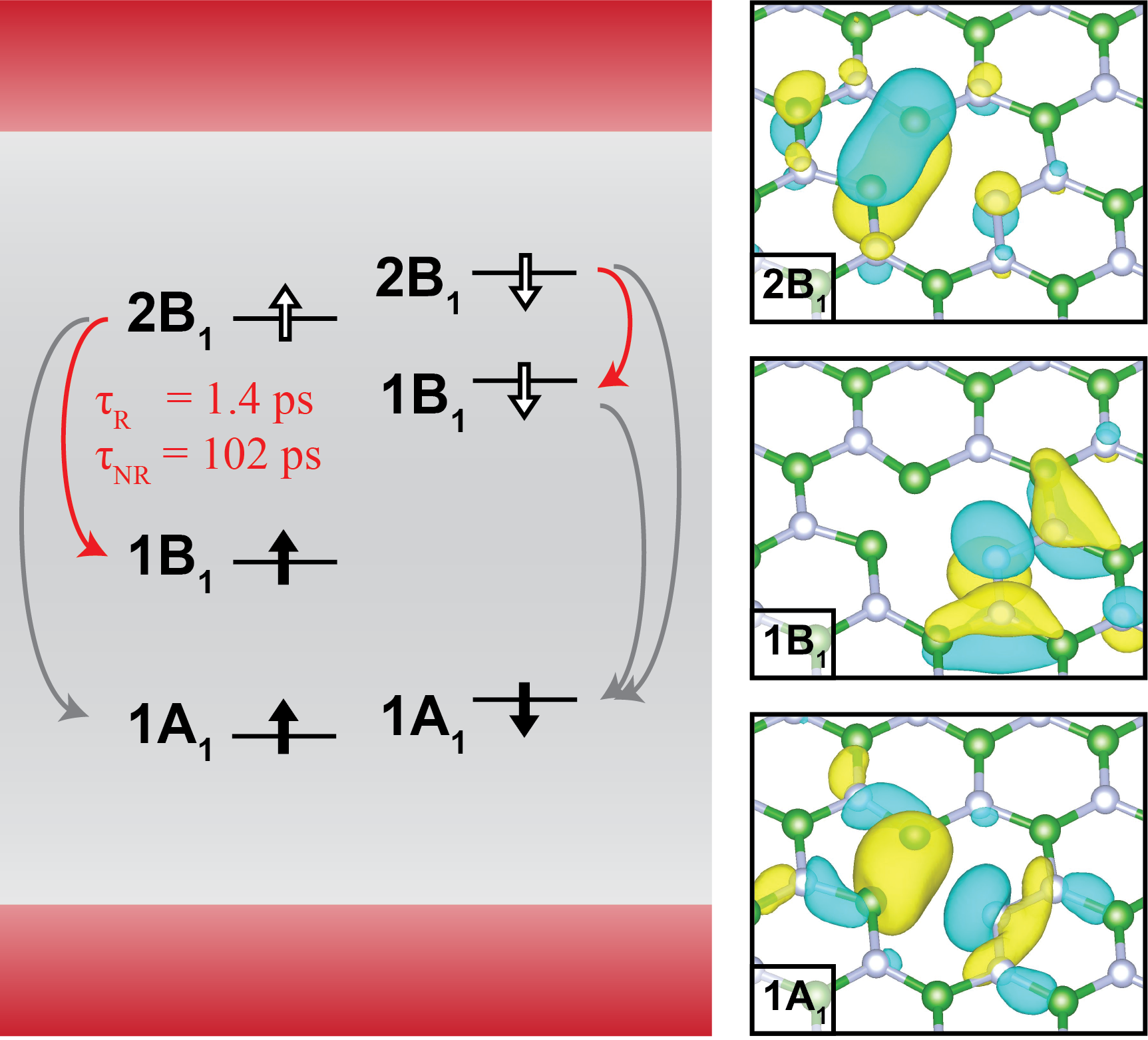}
  \caption{Defect levels and possible defect-defect transitions of \NBVN\ in monolayer $h$-BN. Both up and down spin channels of the $\mathrm{2B_1}/\mathrm{1B_1}$ transitions are marked in red as they are optically allowed with light polarized along defect $C_{2}$ symmetry axis. The exact radiative  ($\uptau_R$) and non-radiative ($\uptau_{NR}$) lifetimes are given for the spin up transition with a 6$\times$6 supercell. The remaining transitions in gray are all  optically forbidden and have very long radiative and non-radiative recombination lifetimes (exceeding 1 ms).
  }
  \label{fig:NBVN-compare-R-NR-all}
\end{figure}

Considering a typical point defect such as \NBVN\, which has been proposed as a promising defect for SPE, \cite{Exarhos2019,Tran2016,Abdi2018} 
we find it introduces several isolated energy levels that lead to multiple possible radiative and non-radiative defect-defect recombination pathways (as shown in Figure \ref{fig:NBVN-compare-R-NR-all}). However, we found only 
the transition between \Bandname{1B1U} and \Bandname{2B1U} (HOMO-LUMO transition for the majority spin channel)
has a viably short radiative lifetime and non-radiative lifetime. All other processes have a non-radiative lifetime longer than ms, much slower than this transition which is at a picosecond level.

\begin{table}[htbp]
    \caption{Properties of defect-defect non-radiative recombination of the \NBVN\ defect in monolayer $h$-BN. Non-radiative lifetimes are computed with a 6$\times$6 supercell at 300 K and $S_{f}$ denotes the ground-state Huang-Rhys factor.}
    \label{table:defect-defect-NBVN}
  \begin{threeparttable}
    \begin{tabular}{ccccccccc}
      \hline
        Transition & \thead{ZPL \\ (eV)} &
        \thead{$\hbar\omega_f$\\ (meV)} &
        \thead{$k$*} &
        \thead{$S_f$} &
        \thead{$X_{if}$}  & 
        \thead{$W_{if}$} & 
        \thead{$C_{p}$ \\ (cm$^2$/s)} & 
        \thead{$\uptau^{NR}$ \\ (ps)} 
        \\
      \hline  
        \Bandname{2B1U} / \Bandname{1B1U} & 2.04 & 100 & 20 &  5.3 
        & 1.3 & 0.38 & $10^{-4}$ & 102 \\
        \Bandname{1B1D} / \Bandname{1A1D} & 1.33 &  58 & 23 & 16.6 
        & $10^{5}$ & $10^{-7}$ & $10^{-11}$ & $>10^{9}$ \\
        \Bandname{2B1D} / \Bandname{1A1D} & 2.94 &  65 & 46 &  7.8 
        & $10^{-4}$ & $10^{-6}$ & $10^{-19}$ & $>10^{9}$ \\
        \Bandname{2B1D} / \Bandname{1B1D} & 1.61 &  57 & 28 &  3.2 
        & $10^{-13}$ & 0.03 & $10^{-19}$ & $>10^{9}$ \\
      \hline
    \end{tabular}
    \begin{tablenotes}
    \item * $k=\Delta E_{if}/\hbar\omega_f$
    \end{tablenotes}
\end{threeparttable}
\end{table}


The non-radiative transition rate is determined by multiple factors based on Eq.~\ref{eq:rate-nonrad-allterms}. 
The first factor is the phonon term $X_{if}$. As the ZPL for all defect-defect transitions are relatively small (less than 3 eV, unlike defect-band transitions), we analyze the subtle difference causing variation of $X_{if}$ among different transitions, based on the relation:
 $X_{if}\propto e^{-S}\frac{S^k}{k!}$ where $k\approx \Delta E_{if}/\hbar\omega_f$ and $S$ is the HR factor.
\cite{Alkauskas2014a}. Specifically ($k>S$ for all defect-defect transitions we study here), a high $S$ implies a large electron-phonon coupling and generally will increase the phonon contribution $X_{if}$.
For example, the HR factor for the \Bandname{1B1D}/\Bandname{1A1D} transition (16.6) is several times larger than other transitions in Table \ref{table:defect-defect-NBVN} and therefore yields the largest  $X_{if}$ of $10^5$ at 300 K. 
On the other hand, a high value of $k$ means a large energy difference (ZPL) relative to the phonon frequency and will reduce phonon contribution $X_{if}$, similar to earlier discussions on defect-band transitions. 
The second factor is the electronic term $W_{if}$, which is proportional to the overlap between electronic wavefunctions $\braket{\psi_i}{\psi_f}$.
Ultimately, only the \Bandname{2B1U}/\Bandname{1B1U} transition has a reasonably large $X_{if}$ and the largest $W_{if}$, which leads to a  viable non-radiative recombination process with a lifetime of 102 ps at 300 K.

The radiative process is more straightforward to interpret as it is directly related to the symmetries of wavefunctions via the dipole transition matrix elements $\mel{\psi_i}{\textbf{r}}{\psi_f}$. 
Computational details can be found in section VI and Table S3 in the SI. The corresponding transition section rules for radiative recombination of \NBVN\ defect in $h$-BN are listed in Table S5.
Both the \Bandname{2B1U}/\Bandname{1B1U} transition and the \Bandname{2B1D}/\Bandname{1B1D} transition are symmetry allowed,\cite{Abdi2018} resulting in short radiative lifetimes of 1.4 ps and 2.5 ps, respectively. This lifetime can be considered to be a lower-bound compared to that of experimental results, because a much higher defect concentration is adopted in practical calculations (1 defect in a 72-atom supercell, i.e. 1 defect per 2 nm$^2$, compared to order of one SPE per $\mu$m$^2$ in experiments \cite{Tran2016}) and both radiative and non-radiative lifetimes will increase linearly with decreasing defect concentrations or increasing supercell size (see Ref.~\citenum{Alkauskas2016} and section V Table S1 and S2 in the SI).
At the low concentration limit, we can consider the defect acts as an isolated molecule in the 2D plane\cite{Wu2019}, 
which gives an upper-bound of the actual lifetime, i.e. 40 ns for \Bandname{2B1U}/\Bandname{1B1U} radiative lifetime at \NBVN. 
This is in good agreement with the experimental radiative lifetime of monolayer $h$-BN SPE, which are measured to be on the order of ns.\cite{Tran2016,Bourrellier2016,Grosso2017,Exarhos2017} We note that different from the recombination rates, the capture coefficient is generally constant as a function of defect concentration or supercell sizes (see Table S1 and S2 in the SI). 

The quantum yield of a SPE (excluding substrate effects) is defined as $\gamma_{if}=r^\mathrm{R}_{if}/(r^\mathrm{R}_{if}+r^{\mathrm{NR}}_{if})$. \cite{Book_Optical_Pankove,Chevallier2017,Hamanaka2014} By comparing the radiative lifetimes with the non-radiative ones shown in Table \ref{table:defect-defect-NBVN} for the \NBVN\ defect, we have $\gamma>98\%$ which shows it has the potential to be a highly efficient quantum emitter. In practice, several other external effects can cause the quantum yield to be substantially lower. In particular, substrate recombination \cite{rosenberger2018electrical}, photobleaching \cite{martinez2016efficient}, and strain (discussed in the next section) are known to play the role of limiting the quantum yield of defect SPE.



\begin{table}[htbp]
    \caption{Properties of $1\mathrm{B}_1\uparrow-2\mathrm{B}_1\uparrow$ defect-defect state transition for \NBVN\ defect in monolayer $h$-BN under strain. Strain directions are shown in Figure \ref{fig:strain-axis}. $X_{if}$ and lifetime are reported at 300 K.    \label{table:defect-defect-NBVN-strain}}
  \begin{threeparttable}
    \begin{tabular}{cccccccc}
      \hline
        Strain &
        \thead{ZPL \\ (eV)} &
        \thead{$\Delta Q$ \\ (\AA) } &
        \thead{$\hbar\omega_f$ \\ (meV)} &  
        \thead{$S_f$} &  
        \thead{$W_{if}$} & 
        \thead{$X_{if}$} & 
        \thead{$\uptau^{NR}$\\ (ps)}  
        \\
      \hline  

No strain &2.04&0.666&100&5.33&0.38&1.26&102 \\
Biaxial $-1\%$&2.08&0.613&105&4.69&0.39&0.28&429 \\
Biaxial $1\%$&2.01&0.732&96&6.18&0.36&9.20&16 \\
Uniaxial $\parallel -1\%$&2.02&0.637&102&4.96&0.39&0.95&127 \\
Uniaxial $\parallel 1\%$&2.07&0.703&98&5.80&0.36&1.98&70  \\
Uniaxial $\bot -1\%$&2.10&0.642&103&5.05&0.38&0.38&336 \\
Uniaxial $\bot 1\%$&1.98&0.697&98&5.69&0.37&5.39&25 \\
      \hline
    \end{tabular}
\end{threeparttable}
\end{table}

\begin{figure}
  \includegraphics[width=0.7\linewidth]{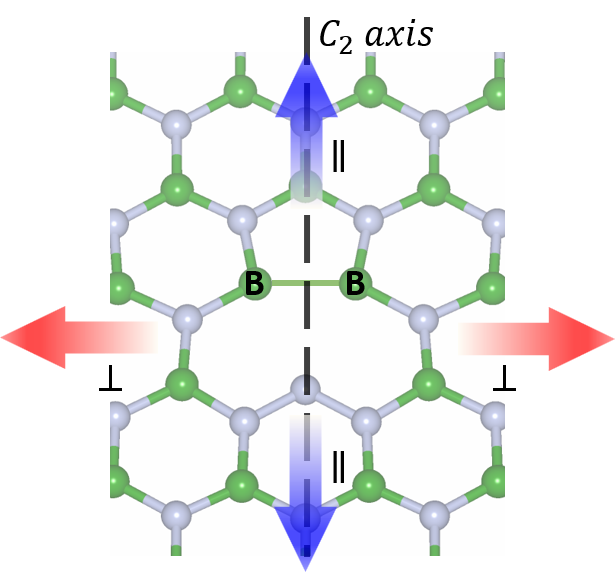}
  \caption{Illustration of the directions of uniaxial strain based on the $C_{2v}$ symmetry of \NBVN\ in $h$-BN. Uniaxial strains applied parallel ($||$ blue arrows) or perpendicular ($\perp$ red arrows) to the $C_{2}$ axis are considered. The optimized atomic structure of \NBVN\ defect is also shown. The green balls denote B atoms and the grey balls denote N atoms.} 
  \label{fig:strain-axis}
\end{figure}

\begin{figure*}
  \includegraphics[width=0.98\linewidth]{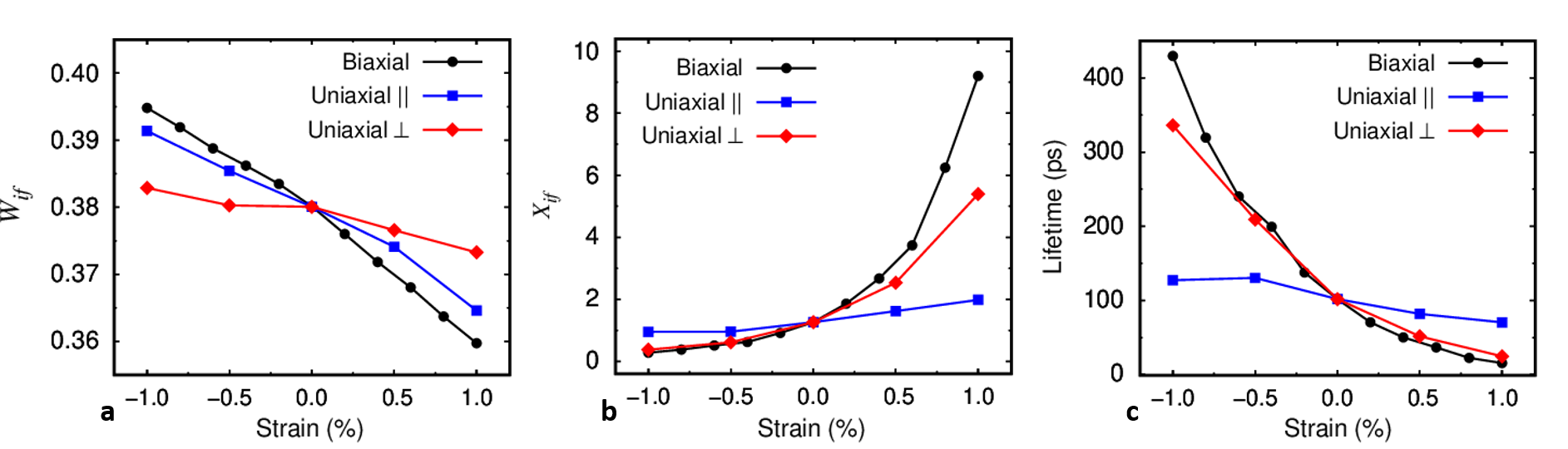}
  \caption{Strain induced properties related to non-radiative recombination lifetime of the $1\mathrm{B}_1-2\mathrm{B}_1$ defect-defect transition of \NBVN\ in monolayer $h$-BN. Strain directions are shown in Figure \ref{fig:strain-axis}.}
  \label{fig:NBVN-strain}
\end{figure*}

In this work, we discuss the impact of strain on non-radiative recombination (and leave other external effects for future study) with \NBVN\ as an example.  
Presumably, strain will change the radiative lifetime little (orbital overlaps between initial and final states are largely preserved) compared to that of the non-radiative lifetime which can be strongly affected by changes in local structures. 
As shown in Figure \ref{fig:strain-axis}, strain may be applied along the $C_{2}$ symmetry axis (denoted as $||$ strain) or  orthogonal to the symmetry axis (denoted as $\perp$ strain). We consider effects of strain along both directions as well as the combinatory effects of biaxial strain for the \Bandname{2B1U}/\Bandname{1B1U} transition (shown in Table \ref{table:defect-defect-NBVN-strain}).   

As discussed earlier, the non-radiative recombination rate is composed of an electronic term $W_{if}$ and a phonon term $X_{if}$. Because $W_{if}$ is proportional to the wavefunction overlap, the change in $W_{if}$ due to strain is found to be negligible, as shown in Figure \ref{fig:NBVN-strain}a.  
However, there are significant changes of the phonon term $X_{if}$ due to strain. 
We note that compressive strain indicates lattice shrinking ($-$); while tensile strain indicates lattice stretching ($+$) and induces opposite changes on non-radiative rates from the former. Therefore, we only discuss compressive strain here. First, compressive strain decreased interatomic distances, which in turn decreased the change in the atomic coordination between initial and final states ($\Delta Q$ in Table~\ref{table:defect-defect-NBVN-strain}). As such, under compressive strain, the HR factor $S=\omega_f\Delta Q^2/2\hbar$ decreased resulting in an exponential decrease of the phonon term $X_{if}$. Such trends occurred regardless of the direction of strain applied (i.e. $\perp$ or $||$ to the $C_2$ axis). Second, a change in the ZPL also occurred under strain \cite{Grosso2017}. After the formation of the nitrogen vacancy, a weak B-B bond is formed perpendicular to the $C_2$ symmetric axis (see Figure \ref{fig:strain-axis}). When compressive strain is applied perpendicular to the $C_2$ axis ($\perp$ strain), the ZPL is increased, due to larger bonding-antibonding splitting of the B-B bond that shifts up the 2B$_1$ energy level (see Figure~\ref{fig:NBVN-compare-R-NR-all} for related wavefunctions and energy levels). As a result, for $\perp$ strain the change in ZPL and HR factor coincided and yielded an exponential decrease of $X_{if}$ under compressive strain (red curve Figure \ref{fig:NBVN-strain}b). In contrast, for $||$ strain, these changes counteracted each other resulting in a nearly constant value of $X_{if}$ (blue curve Figure \ref{fig:NBVN-strain}b). In addition, biaxial strain is a simple combinatory effect of $||$ and $\perp$ strain, mostly dominated by the trend of $\perp$ strain (black curve Figure \ref{fig:NBVN-strain}b).
All in all, the exponential change in $X_{if}$ for both biaxial and uniaxial $\perp$ strain resulted in exponential modification to the non-radiative lifetime of the defect, as black and red curves shown in Figure \ref{fig:NBVN-strain}c. 
In particular, in the case of tensile biaxial strain, with $1\%$, the quantum yield decreased by $10\%$ due to an order of magnitude decrease in non-radiative lifetime. This highlights the significant impact strain can have on the efficiency of defect SPE.


In summary, in this work we compared the radiative and phonon-assisted non-radiative recombinations at defects in wide bandgap 2D materials, using monolayer $h$-BN as a prototypical example. We found the radiative recombination rates far surpass the non-radiative ones, highlighting the potential of point defects in wide bandgap 2D materials as single photon emitters. Defect-band non-radiative recombinations all have negligible rates possibly due to large energy differences between initial and final states, and only a small subset of defect-defect non-radiative transitions are possible. Transitions vary on several orders of magnitude due to wavefunction symmetry, HR factor, as well as zero-phonon line (ZPL). Finally, we show that compressive or tensile strain up to 1\% can alter the non-radiative lifetime by orders of magnitude. 
Hence, strain largely impacts the quantum yield of single photon emitters and alters the photon energy of the emitter for use towards specific optoelectronic applications. Our study provides important insights on the critical factors of defects in 2D materials as single photon emitters for quantum information applications.

We acknowledge Lin-Wang Wang, Audrius Alkauskas and Cyrus Dreyer for helpful discussions. This work is supported by National Science Foundation under grant No. DMR-1760260 and DMR-1747426, and the Hellman Fellowship. This research used resources of the Center for Functional Nanomaterials, which is a US DOE Office of Science Facility, and the Scientific Data and Computing center, a component of the Computational Science Initiative, at Brookhaven National Laboratory under Contract No. DE-SC0012704, the National Energy Research Scientific Computing Center (NERSC) a U.S. Department of Energy Office of Science User Facility operated under Contract No. DE-AC02-05CH11231, the Extreme Science and Engineering Discovery Environment (XSEDE) which is supported by National Science Foundation Grant No. ACI-1548562 \cite{xsede}. 



\end{document}